\documentstyle[prl,aps,twocolumn]{revtex}
\begin{document}
\draft

\twocolumn[\hsize\textwidth\columnwidth\hsize\csname @twocolumnfalse\endcsname
\title{Matter Mass Generation and $\Theta$ Vacuum: Dynamical Spontaneous Symmetry Breaking}
\author{Heui-Seol Roh\thanks{e-mail: hroh@nature.skku.ac.kr}}
\address{BK21 Physics Research Division, Department of Physics, Sung Kyun Kwan University, Suwon 440-746, Republic of Korea}
\date{\today}
\maketitle

\begin{abstract}
This work proposes a stringent concept of matter mass generation and $\Theta$ vacuum
in the context of local gauge theory for the strong force under the constraint of the
flat universe. The matter mass is generated as the consequence of dynamical
spontaneous symmetry breaking (DSSB) of gauge symmetry and discrete symmetries, which
is motivated by the parameter $\Theta$ representing the surface term. Matter mass
generation introduces the typical features of constituent particle mass, dual Meissner
effect, and hyperfine structure. The $\Theta$ term plays important roles on the DSSB
of the gauge group and on the quantization of the matter and vacuum space. The
$\Theta$ vacuum exhibits the intrinsic principal number and intrinsic angular momentum
for intrinsic space quantization in analogy with the extrinsic principal number and
extrinsic angular momentum for extrinsic space quantization.
\end{abstract}

]
\narrowtext

The understanding of matter mass generation and $\Theta$ vacuum \cite{Hoof2} is one of
the longstanding problems in physics, which are deeply related to underlying
principles of the universe. Quantum chromodynamics (QCD) \cite{Frit} does not
perturbatively provide any clues for matter mass generation. Although the
Glashow-Weinberg-Salam (GWS) model \cite{Glas} holds many attractive features, the
Higgs mechanism \cite{Higg} and fermion mass generation are the least satisfied
aspects of electroweak interactions. The Higgs mechanism in the GWS model is applied
to generate the masses both of gauge bosons and of fermions, but the masses of
fermions are just input parameters of the theory which are not predicted. The
difficulty comes from the fact that conventional mass terms in the Lagrangian are not
allowed because the left- and right-handed components of the various elementary
fermion fields have different quantum numbers as seen in the GWS model \cite{Glas} and
so simple mass terms violate gauge invariance. The Higgs mechanism or spontaneous
symmetry breaking mechanism fails to generate fermion masses even though it is
successful in generating gauge boson masses: experimentally, Higgs particles are not
observed yet. In this scheme, the $\Theta$ vacuum term as the nonperturbative one is
taken into account to show dynamical spontaneous symmetry breaking (DSSB), which is
successful in the generation of fermion masses as well as gauge boson masses and is
relevant with the axial current anomaly in strong interactions \cite{Adle,Hoof2,Roh3}.
This approach thus concentrates on the DSSB mechanism of local gauge symmetry and
global discrete symmetries to generate fermion masses. The $\Theta$ term representing
the surface effect takes the difference in quantum numbers of left- and right-handed
fermions to generate elementary fermion masses. Due to the $\Theta$ term, the boundary
condition of the system is imposed and the matter-vacuum space is quantized. Not only
mass generation mechanism but also $\Theta$ vacuum may be resolved systematically.
This paper briefly addresses characteristics of mass generation and $\Theta$ vacuum in
strong interactions, whose details are addressed in the reference \cite{Roh3}.

The nonperturbative $\Theta$ vacuum term \cite{Hoof2} is added to the perturbative gauge theory in order to generate
fermion and gauge boson mass terms nonperturbatively.
The QCD Lagrangian density \cite{Roh3} is given by
\begin{equation}
{\cal L}_{QCD} = - \frac{1}{2} Tr  G_{\mu \nu} G^{\mu \nu}
+ \sum_{i=1}  \bar \psi_i i \gamma^\mu D_\mu \psi_i  + \Theta \frac{g_s^2}{16 \pi^2} Tr G^{\mu \nu} \tilde G_{\mu \nu},
\end{equation}
where the bare $\Theta$ term \cite{Hoof2} is a nonperturbative term added to the
perturbative Lagrangian density with an $SU(3)_C$ gauge invariance. The subscript $i$
stands for the classes of pointlike spinor $\psi$ and $A_{\mu} = \sum_{a=0} A^a_{\mu}
\lambda^a /2$ stand for gauge fields. The field strength tensor is given by $G_{\mu
\nu} = \partial_\mu A_\nu -
\partial_\nu A_\mu - i g_s [A_\mu, A_\nu]$ and $\tilde G_{\mu
\nu}$ is the dual field strength tensor. The $\Theta$ term
apparently odd under both P and T operation plays the role
relating two different worlds, matter and vacuum.
Since conventional mass terms in
the Lagrangian density are not allowed, this scheme adopts the
DSSB of local gauge symmetry as well as global chiral symmetry
\cite{Namb}, which uses the fermion condensation as the order
parameter, to generate fermion masses. The Lagrangian density is
invariant under global gauge transformation but the true vacuum is
not invariant due to the condensation of singlet gauge bosons and
fermions.
The gauge boson mass $M_G$ decreases through
the condensation of the singlet gauge boson:
\begin{equation}
M_G^2 = M_{H}^2 - c_f g_s^2 \langle \phi \rangle^2 = c_f g_s^2 [A_{0}^2 -
\langle  \phi \rangle^2]
\end{equation}
where $M_{H} = \sqrt{c_f} g_s A_{0}$ is the gauge boson mass at the grand unification
scale, $A_{0}$ is the singlet gauge boson with even parity, and $\langle  \phi \rangle$
is the condensation of singlet gauge boson with odd parity. The effective coupling
constant at the strong scale is expressed in analogy with the phenomenological,
electroweak coupling constant $G_F =  \frac{\sqrt{2} g_w^2}{8 M_G^2}$ with the weak
coupling constant $g_w$:
\begin{equation}
\frac{G_R}{\sqrt{2}} = - \frac{c_f g_s^2}{8 (k^2 - M_G^2)} \simeq \frac{c_f g_s^2}{8
M_G^2}
\end{equation}
where $k$ denotes the four momentum and $c_f$ denotes the color factor. DSSB consists
of two simultaneous mechanisms; the first mechanism is the explicit symmetry breaking
of gauge fields, which is represented by the strong coupling constant $c_f \alpha_s$
with the color factor $c_f$ and the strong fine structure constant $\alpha_s$, and the
second mechanism is the spontaneous symmetry breaking of gauge fields, which is
represented by the condensation of singlet gauge fields. A metastable, unphysical
vacuum leads to a stable physical vacuum through DSSB due to the condensation of
singlet gauge bosons. The condensation reducing the vacuum energy generates fermion
masses as a consequence of gauge and discrete symmetries breaking in course of parity
and charge conjugation violation. For strong interactions, the $SU(2)_V$ doublet
vector current is conserved but the $SU(2)_A$ (or $U(1)_A$) singlet axial vector
current is not conserved during DSSB and for electroweak interactions, the $SU(2)_L$
doublet (V - A) current is conserved but the $SU(2)_R$ (or $U(1)_R$) singlet (V + A)
current is not conserved during DSSB \cite{Glas,Roh3}.
The proton and neutron as spinors possess up and down colorspins as a
doublet just like up and down strong isospins:
\begin{equation}
{\uparrow \choose \downarrow}_c, \ \uparrow = {1 \choose 0}_c, \ \downarrow = {0
\choose 1}_c .
\end{equation}
This implies that conventional, global $SU(2)$ strong isospin symmetry introduced by Heisenberg \cite{Heis}
is postulated as the combination of local $SU(2)$ colorspin and local $SU(2)$ weak isospin symmetries.

The constraint of the extremely flat universe
\begin{equation}
\Omega -1 = - 10^{-61} ,
\end{equation}
which is required by quantum gauge theory, provides the relation $\Omega = (\langle
\rho_m \rangle - \Theta \rho_m)/\rho_G$ where $\langle \rho_m \rangle$ is the zero
point energy density, $\rho_m$ is the matter energy density, $\rho_G = M_G^4$ is the
vacuum energy density, and
\begin{equation}
\Theta = 10^{-60} \rho_G/\rho_m
\end{equation}
is the parameter of the $\Theta$ vacuum term representing the surface term
\cite{Roh3}. Matter mass generation has features represented by the $\Theta$ vacuum,
dual Meissner effect, and constituent particle mass \cite{Roh3}. The matter mass is
attributed from the dual pairing process due to dielectric mechanism, which violates
the gauge symmetry and discrete symmetries. The relation between the gauge boson mass
and the fermion mass is given by
\begin{equation}
M_G = \sqrt{\pi} m_f c_f \alpha_s \sqrt{N_{sd}}
\end{equation}
where $N_{sd}$ is the difference
number of even-odd parity singlet fermions in intrinsic two-space dimensions. The
above relation stems from the dual pairing mechanism
\begin{math}
M_G = g_{sm}^2 |\psi (0)|^2/ m_f ,
\end{math}
in analogy with electric superconductivity, where $|\psi (0)|^2 = (m_f c_f
\alpha_s)^3$ is the particle probability density and $g_{sm} = 2 \pi n/\sqrt{c_f} g_s
= 2 \pi \sqrt{N_{sd}}/\sqrt{c_f} g_s$ is the color magnetic coupling constant
according to the Dirac quantization condition \cite{Dira}:
\begin{equation}
\sqrt{c_f} g_s g_{sm} = 2 \pi \sqrt{N_{sd}} .
\end{equation}
The particle number
$N_{sd}$ is the difference number between the number $N_{ss}$ of singlet particles
interacting with charge symmetric configurations and the number $N_{sc}$ of condensed
particles interacting with charge asymmetric configurations: $N_{sd} = N_{ss} -
N_{sc}$. The fermion mass formed as the result of confinement mechanism is composed of
constituent particles:
\begin{math}
m_f = \sum_i^N m_i
\end{math}
where $m_i$ is the constituent particle mass as the result of the
$U(1)$ gauge theory. In the above, $N$ depends on the intrinsic
quantum number of constituent particles: $N = N_{sd}^{3/2}$.
For examples, $N = 1/B$
with the baryon number $B$ for the constituent quark in the
formation of a baryon, $N = 1/M$ with the meson number $M$ for the
constituent quark in the formation of a meson, and $N = 1/L$ with
the lepton number $L$ for a constituent particle in the formation
of a lepton. A fermion mass term in the Dirac Lagrangian has the
form $m_f \bar \psi \psi = m_f (\bar \psi_A \psi_V + \bar \psi_V
\psi_A)$ where the mass term is equivalent to a helicity flip.
Vecor fermions are put into $SU(2)$ doublets and
axial-vector ones into $SU(2)$ singlets. In the matter space, it is the
pairing mechanism of electric monopoles while in the vacuum space, it is
the pairing mechanism of magnetic monopoles according to the
duality between electricity and magnetism \cite{Dira}. In the
dual pairing mechanism, discrete symmetries P, C, T, and CP are
dynamically broken due to massive gauge bosons \cite{Roh3}.
Electric monopole, magnetic dipole, and
electric quadrupole remain in the matter space but magnetic monopole,
electric dipole, and magnetic quadrupole condense in the vacuum space as
the consequence of P violation. Antibaryon particles condense in the
vacuum space while baryon particles remain in the matter space as the result of
C and CP violation at the strong scale: the baryon
asymmetry $\delta_B \simeq 10^{-10}$ \cite{Stei0}. The electric dipole moment of the neutron
and no parity partners in hadron spectra are the typical examples for P, T, and CP
violation at the strong scale. The fine or hyperfine
structures of a fermion mass include spin-spin, isospin-isospin,
and colorspin-colorspin interactions due to intrinsic angular
momenta of $SU(2)$ gauge theories:
\begin{equation}
\triangle E \propto \alpha_m \frac{\vec s_i \cdot \vec s_j}{m_i m_j} |\psi (0)|^2
+ \alpha_i \frac{\vec i_i \cdot \vec i_j}{m_i m_j} |\psi (0)|^2
+ \alpha_s \frac{\vec \zeta_i \cdot \vec \zeta_j}{m_i m_j} |\psi (0)|^2
\end{equation}
where $\alpha_m$, $\alpha_i$, and $\alpha_s$ are respectively
coupling constants. The approach can be applied to investigate the masses of hadrons,
which justify the constituent quark model \cite{Dash} as an effective model of
QCD at low energies from this viewpoint. If fine structure and
hyperfine structure interactions due to colorspin and isospin
contributions are absorbed to the constituent quark mass, the
meson mass of the conventional constituent quark model is
obtained:
\begin{equation}
m_m = m_1 + m_2 + A \frac{\vec \sigma_1 \cdot \vec \sigma_2}{m_1 m_2}
\end{equation}
where $\vec \sigma_1 \cdot \vec \sigma_2 = 4 \vec s_1 \cdot \vec s_2 =1$ for vector mesons and
$\vec \sigma_1 \cdot \vec \sigma_2 = -3$ for pseudoscalar mesons are given
and $A = \frac{8 \pi g_s^2 |\psi (0)|^2}{9}$.
By the same token, the baryon mass of the conventional constituent quark model is obtained:
\begin{equation}
m_b = m_1 + m_2 + m_3 + A' \sum_{i>j} \frac{\vec \sigma_i \cdot \vec \sigma_j}{m_i m_j}
\end{equation}
where $A' = \frac{4 \pi g_s^2 |\psi (0)|^2}{9}$.
Since $\sum \sigma_i \cdot \sigma_j = 4 s_i \cdot s_j = 2 [s(s+1) - 3s(s+1)]$ with the total spin
$\vec S = \vec s_1 + \vec s_2 + \vec s_3$,
$\sum \vec \sigma_i \cdot \vec \sigma_j = 3$ for decuplet baryons and
$\sum \vec \sigma_i \cdot \vec \sigma_j = -3$ for octet baryons are given.
The constituent quark model illustrates reasonable agreement in hadron spectra within a few percent deviation.

The difference number $N_{sd}$ in intrinsic two-space dimensions suggests a
degenerated particle number $N_{sp}$ and an intrinsic principal number $n_m$ in the
intrinsic radial coordinate; these quantum numbers are connected by the relation
$n_m^4 = N_{sp}^2 = N_{sd}$ and the Dirac quantization condition $\sqrt{c_f} g_s
g_{sm} = 2 \pi N_{sp}$ is satisfied. The $N_{sp}$ is thus the degenerated state number
in the intrinsic radial coordinate that has the same principal number $n_m$. The
intrinsic principal quantum number $n_m$ consists of three quantum numbers, that is,
$n_m = (n_c, n_i, n_s)$ where $n_c$ is the intrinsic principal quantum number for the
color space, $n_i$ is one for the isospin space, $n_s$ is one for the spin space.
Intrinsic quantum numbers $(n_c, n_i, n_s)$ take integer numbers. A fermion therefore
possesses a set of intrinsic quantum numbers $(n_c, n_i, n_s)$ to represent its
intrinsic quantum state. The concept automatically adopts the three types of intrinsic
angular momentum operators, $\hat C$, $\hat I$, and $\hat S$, when intrinsic
potentials for color, isospin, and spin charges are central so that they depend on the
internal radial distance: for instance, the color potential in strong interactions is
dependent on the radial distance. The intrinsic spin operator $\hat S$ has a magnitude
square $\langle S^2 \rangle = s (s + 1)$ and $s = 0, 1/2, 1, 3/2 \cdot \cdot \cdot
(n_s-1)$: the third component of $\hat S$, $\hat S_z$, takes half integer or integer
quantum number in the range of $- s \sim s$ with the degeneracy $2s + 1$. The
intrinsic isospin operator $\hat I$ and intrinsic color operator $\hat C$ are
analogously quantized. The principal number $n_m$ in intrinsic space quantization is
very much analogous to the principal number $n$ in extrinsic space quantization and
the intrinsic angular momenta are analogous to the extrinsic angular momentum so that
the total angular momentum has the form of
\begin{equation}
\vec J = \vec L + \vec S + \vec I + \vec C  ,
\end{equation}
which is the extension of the conventional total angular momentum $\vec J = \vec L +
\vec S$. The intrinsic principal number $n_m = (n_c, n_i, n_s)$ denotes the intrinsic
spatial dimension or radial quantization: $n_c = 3$ represents strong interactions as
an $SU(3)_C$ gauge theory. For QCD as the $SU(3)_C$ gauge theory, there are nine
gluons ($n_c^2 = 3^2$), which consist of one singlet gluon $A_0$ with $c=0$, three
degenerate gluons $A_1 \sim A_3$ with $c=1$, and five degenerate gluons $A_4 \sim A_8$
with $c=2$. One explicit evidence of colorspin and isospin angular momenta is strong
isospin symmetry in nucleons, which is postulated as the combination symmetry of
colorspin and weak isospin. Another evidence is the nuclear magnetic dipole moment:
the Lande spin g-factors of the proton and neutron are respectively $g_s^p = 5.59$ and
$g_s^n = - 3.83$ shifted from $2$ and $0$ because of contributions from color and
isospin as well as spin. The mass ratio of the proton and the constituent quark,
$m_p/m_q \sim 2.79$, thus represents three degrees of freedom of color, isospin, and
spin. In fact, the extrinsic angular momentum associated with the intrinsic angular
momentum may be decomposed by
\begin{equation}
\vec L = \vec L_i + \vec L_c + \vec L_s
\end{equation}
where $\vec L_i$, $\vec L_c$, and $\vec L_s$ are angular momenta originated from
isospin, colorspin, and spin, respectively. This is supported by the fact that the
orbital angular momentum $l_c$ of the nucleon has the different origin from the color
charge with the orbital angular momentum $l_i$ of the electron from the isospin charge
since two angular momenta have opposite directions from the information of spin-orbit
couplings in nucleus and atoms. Extrinsic angular momenta have extrinsic parity
$(-1)^l = (-1)^{(l_c + l_i + l_s)}$, intrinsic angular momenta have intrinsic parity
$(-1)^{(c + i + s)}$, and the total parity becomes $(-1)^{(l + c + i + s)}$ for
electric moments while extrinsic angular momenta have opposite parities respectively.
Fermions increase their masses by decreasing their intrinsic principal quantum numbers
from the higher ones at higher energies to the lower ones at lower energies. The
coupling constant $\alpha_s$ of a non-Abelian gauge theory is strong for the small
$N_{sd}$ and is weak for the large $N_{sd}$ according to the renormalization group
analysis. The vacuum energy is described by the zero-point energy in the unit of
$\omega/2$ with the maximum number $N_{sd} \simeq 10^{61}$ and the vacuum is filled
with fermion pairs of up and down colorspins, isospins, or spins, whose pairs behave
like bosons quantized by the unit of $\omega$: this is analogous to the
superconducting state of fermion pairs. The intrinsic particle number $N_{sp} \simeq
1$ (or $B \simeq 1$) characterizes strong interactions for nucleons: according to $M_G
= \sqrt{\pi} m_h c_f \alpha_s \sqrt{N_{sd}}$ \cite{Roh3}, $m_h = 0.94$ GeV, $c_f =
1/3$, $\alpha_s = 0.48$, $N_{sd} = 1$, $M_G = 0.27$ GeV for a nucleon are realized.
Note that if $N_{sp}
> 1$ (or $B < 1$), it represents a pointlike fermion and if $N_{sp} < 1$ (or $B > 1$),
it represents a composite fermion.

The invariance of gauge transformation provides
$\psi [\hat O_\nu] = e^{i \nu \Theta} \psi [\hat O]$ for the fermion wave function $\psi$ with the transformation of an operator $\hat O$ by
the class $\nu$ gauge transformation, $\hat O_\nu$:
the vacuum state characterized by the constant $\Theta$ is called the $\Theta$ vacuum \cite{Hoof2}.
The true vacuum is the superposition of all the $|\nu \rangle$ vacua with the phase $e^{i \nu \Theta}$:
$|\Theta \rangle = \sum_\nu e^{i \nu \Theta} |\nu \rangle$.
The topological winding number $\nu$ or the topological charge $q_s$ is defined by
\begin{equation}
\nu = \nu_+ - \nu_- = \int \frac{c_f g_s^2}{16 \pi^2} Tr G^{\mu \nu} \tilde G_{\mu \nu} d^4 x
\end{equation}
where the subscripts $+$ and $-$ denote moving particles with opposite
intrinsic properties in the presence of the gauge fields \cite{Atiy}.
The subscripts $+$ and $-$ represents axial-vector and vector particles at the strong scale.
The matter energy density generated by the surface effect is postulated by
\begin{equation}
\rho_m \simeq \rho_c \simeq \frac{c_f g_s^2}{16 \pi^2} Tr G^{\mu
\nu} \tilde G_{\mu \nu}  \simeq 10^{-47} \ \textup{GeV}^4
\end{equation}
which implies that the fermion mass is generated by the difference of fermion numbers
moving to opposite directions. The difference number $N_{sd}$, the singlet fermion
number $N_{ss}$, and the condensed singlet fermion number $N_{sc}$ in intrinsic
two-space dimensions respectively correspond to $\nu$, $\nu_+$, and $\nu_-$ in
three-space and one-time dimensions. In the presence of the $\Theta$ term, the axial
vector current is not conserved due to an Adler-Bell-Jackiw anomaly \cite{Adle}:
\begin{equation}
\partial_\mu J_\mu^5 = \frac{N_f c_f g_s^2}{16 \pi^2} Tr G^{\mu \nu} \tilde G_{\mu \nu}
\end{equation}
with the class number of fermions $N_f$ and this reflects
degenerated multiple vacua. This illustrates mass generation by
the surface effect due to the field configurations with parallel charge
electric and magnetic fields. $\Theta$ values defined by $\Theta =
10^{-61} \rho_G/\rho_m$ is consistent with the observed results,
$\Theta < 10^{-9}$ in the electric dipole moment of the neutron \cite{Alta}.
The topological winding number parameterized by $\nu = \rho_m/\rho_G$ is related to
the intrinsic quantum number $n_m$ by $\nu = 1/n_m^{8}$.
The intrinsic principal number $n_m$ is also connected with
$N_{sp}$ and $N_{sd}$: $n_m^2 = N_{sp}$, $N_{sp}^2 = N_{sd}$, and
$N_{sp}^{4} = 1/\nu$. The relation between the intrinsic
radius and the intrinsic quantum number might be ascribed by $r_i
= r_{0i} / n_m^2$ with the radius $r_{0i} \simeq 1/m_f c_f
\alpha_s \simeq N_{sp}/M_G$. Intrinsic quantum numbers are exactly
analogous to extrinsic quantum numbers. The extrinsic principal
number $n$ for the nucleon is related to the nuclear mass number
$A$: $n^2 = A^{1/3}$, $n^4 = A^{2/3}$, $n^6 = B = A$. The relation
between the nuclear radius and extrinsic quantum number is
outlined by
\begin{equation}
r = r_0 A^{1/3} = r_0 n^2
\end{equation}
with the radius $r_0 \approx 1.2$ fm. This is analogous to the atomic radius $r_e =
r_0 n_e^2$ with the atomic radius $r_0$ and the electric principal number $n_e$: the
atomic radius $r_0$ is almost the same with the Bohr radius $a_B = 1/m_e \alpha_e =
0.5 \times 10^{-8}$ cm. These concepts are related to the constant nuclear density
$n_B = 3/4 \pi r_0^3 = 1.95 \times 10^{38} \ \textup{cm}^{-3}$ or Avogadro's number
$N_A = 6.02 \times 10^{23} \ \textup{mol}^{-1}$ and to the constant electron density
$n_e = 3/4 \pi r_s^3 = 6.02 \times 10^{23} Z \rho_B/A$ with the matter energy density
$\rho_m$ in $g/\textup{cm}^3$ where the possible relation is
\begin{math}
r_e = r_0 L^{1/3} = r_0 n_e^2
\end{math}
with the lepton number $L$. The $\Theta$ value indicating DSSB becomes
$\Theta_{QCD} \approx 10^{-12}$ at the strong scale.
Baryon mass generation is described by $\rho_B \equiv
\Omega_B \rho_c \simeq 10^{-61} \Omega_B \rho_G/\Theta \simeq \Omega_B 10^{-47}
\ \textup{GeV}^4$ with the gluon mass density $\rho_G \approx
10^{-2} \ \textup{GeV}^4$ at the strong scale. $\Theta$ term as
the surface term modifies the original QCD for strong interactions
\cite{Frit}, which has the problem in fermion mass violating
gauge invariance, and suggest mass generation as the
nonperturbative breaking of gauge and chiral invariance through
DSSB. The presence of the nonperturbative anomaly represented by
the $\Theta$ vacuum does not necessarily spoil renormalization
because there is no Ward-Takahashi identity \cite{Ward} destroyed
by the nonconservation of any local current with even parity.

New development based on a local gauge theory for the strong force under the
constraint of the flat universe $\Omega - 1 = - 10^{-61}$ are matter mass generation
mechanism through the dynamical spontaneous breaking of gauge symmetries and $\Theta$
vacuum structure through intrinsic quantization specifying intrinsic properties.
Matter mass generation introduces the typical features of constituent particle mass induced from
the gauge boson mass, dual Meissner effect through the pairing mechanism, and fine
structure due to intrinsic angular momentum interactions. The $\Theta$ term plays
important roles on the DSSB of the gauge group and on the quantization of the matter
and vacuum space. $\Theta$ vacuum structure exhibits intrinsic principal numbers and
angular momenta for intrinsic space quantization in analogy with the extrinsic
principal number and angular momentum for extrinsic space quantization. The intrinsic
space essentially consists of the color space, isospin space, and spin space, which
form their associated extrinsic space: the total angular momentum $\vec J = \vec C +
\vec I + \vec S + \vec L$ is the extension of the conventional one $\vec J = \vec S +
\vec L$. This work may thus become a turning point toward the complete understanding
of the universe structure since the insight in matter mass generation and $\Theta$
vacuum is clarified.

\end{document}